\documentclass[a4paper,fleqn]{cas-dc}
\usepackage{textcomp}
\usepackage{stfloats}
\usepackage{url}
\usepackage{verbatim}
\usepackage{graphicx}
\usepackage{cite}
\usepackage{multirow,booktabs,amssymb}
\usepackage{pifont}
\usepackage{glossaries}
\usepackage{color}
\usepackage[authoryear,longnamesfirst]{natbib}
\newacronym{fsd}{FSD}{fake speech detection}

\def\tsc#1{\csdef{#1}{\textsc{\lowercase{#1}}\xspace}}
\tsc{WGM}
\tsc{QE}
\tsc{EP}
\tsc{PMS}
\tsc{BEC}
\tsc{DE}

\begin{document}
	\let\WriteBookmarks\relax
	\def\floatpagepagefraction{1}
	\def\textpagefraction{.001}
	\shorttitle{Spatial Reconstructed Local Attention Res2Net with F0 Subband for Fake Speech Detection}
	
	\shortauthors{Cunhang Fan et~al.}
	
	\title [mode = title]{Spatial Reconstructed Local Attention Res2Net with F0 Subband for Fake Speech Detection}                      
	\tnotemark[1]
	


	%
	\author[1]{Cunhang Fan}
	
	
	\affiliation[1]{organization={Anhui Province Key Laboratory of Multimodal Cognitive Computation, School of Computer Science and Technology, Anhui University},
		city={Hefei},
		postcode={230601}, 
		country={China}}
	
	\author[1]{Jun Xue}
	\author[2]{Jianhua Tao}
	
	\affiliation[2]{organization={Department of Automation, Tsinghua University},
		city={Beijing},
		postcode={100190}, 
		country={China}}
	
	\author[3]{Jiangyan Yi}
	\affiliation[3]{organization={National Laboratory of Pattern Recognition, Institute of Automation, Chinese Academy of Sciences},
		city={Beijing},
		postcode={100190}, 
		country={China}}
	\author[3]{Chenglong Wang} 	
	
	\author[4]{Chengshi Zheng}

	\affiliation[4]{organization={Key Laboratory of Noise and Vibration Research, Institute of Acoustics, Chinese Academy of Sciences},	city={Beijing},
		postcode={100190}, 
		country={China}}

	\author[1]{Zhao Lv}
	\ead{kjlz@ahu.edu.cn}
	
	\author[2]{Jianhua Tao}
\ead{jhtao@tsinghua.edu.cn}
	
	\cortext[cor1]{Corresponding author: Zhao Lv and Jianhua Tao}
	
	%

	\begin{abstract}
		The rhythm of bonafide speech is often difficult to replicate, which causes that the fundamental frequency (F0) of synthetic speech is significantly different from that of real speech.
		It is expected that the F0 feature contains the discriminative information for the  \acrfull{fsd} task. 
		In this paper, we propose a novel F0 subband for \acrshort{fsd}. In addition, to effectively model the F0 subband so as to improve the performance of \acrshort{fsd}, the spatial reconstructed local attention Res2Net (SR-LA Res2Net) is proposed. Specifically, Res2Net is used as a backbone network to obtain multiscale information, and enhanced with a spatial reconstruction mechanism to avoid losing important information when the channel group is constantly superimposed.
		In addition, local attention is designed to make the model focus on the local information of the F0 subband. Experimental results on the ASVspoof 2019 LA dataset show that our proposed method obtains an equal error rate (EER) of 0.47\% and a minimum tandem detection cost function (min t-DCF) of 0.0159, achieving the state-of-the-art performance among all of the single systems. 
	\end{abstract}
	
	
	
	
		\maketitle

	\section{Introduction}
	
	Automatic speaker verification (ASV) technology has become increasingly mature, but it remains vulnerable to attack by existing synthetic speech techniques. Generally, fake speech can be divided into three types: audio playback~\cite{fan2023two,shang2008preliminary, paul2016countermeasure, kinnunen2017reddots, Tomietal, fan2023compnet,hajipour2021listening, ali2021fake}, text-to-speech (TTS)~\cite{Shchemelinin-et-al:scheme, huang2022meta, zhang2021fmfcc}, and voice conversion (VC)~\cite{kinnunen-et-al:scheme, chen2020generalization, tian2017exemplar}. To reduce the risk of spoofing attacks on ASV caused by fake audio, the ASVspoof challenges have been held successively in 2015~\cite{Wu-et-al:scheme}, 2017~\cite{Tomietal}, 2019~\cite{Todiscoetal}, and 2021~\cite{Yamagishi}. In 2022, the Audio Deep Synthesis Detection (ADD 2022)~\cite{yi2022add} was also successfully held.
		The ASVspoof challenge has two sub-challenges, one is logical access  (LA)\footnote{Proposed by the ASVspoof challenge, it is a scenario for the study of fake speech detection.} attacks using TTS and VC algorithms, and the other is physical access (PA) attacks using audio playback. The research in this paper focuses on LA attacks.
	Currently, the main focus of research in \acrfull{fsd} lies in the design of front-end features and back-end models.
	
	For front-end features, many acoustic features are investigated~\cite{doan2023bts,huang2023discriminative,das2019long,wei2022new,li2022long,williams2019speech,yang2020long,paul2017spectral,yang2019extraction,fan2023subband}, such as Mel Frequency Cepstral Coefficients (MFCC), constant Q cepstral coefficients (CQCC), linear frequency cepstral coefficients (LFCC) and so on. In addition,  in~\cite{witkowski-et-al:scheme}, it is proposed to use Inverse MFCC (IMFCC), Linear Prediction Cepstral Coefficients (LPCC), and LPCCres\footnote{Proposed by the authors of \cite{witkowski-et-al:scheme}, and derived by the LPCC.} features. 
	Then, the high-frequency components of these three features are fed to a classifier that classifies real samples and replayed samples.
	In~\cite{Bhusan-et-al:scheme}, it is proposed to divide the whole frequency band into multiple disjoint sub-bands.  
		A joint subband modeling architecture is designed to learn the specific features of subbands.
	In~\cite{Zhang-et-al:scheme}, it is proposed to divide the log power spectrogram (LPS) feature frequency band into two frequency bands, namely high frequency and low frequency. Based on the results of the experiments, low frequency has superior performance to high frequency.
	While these methods make significant advances in \acrshort{fsd} and demonstrate that different frequency bands have different effects, they do not specify how the bands should be divided.

	For back-end models, deep neural network (DNN)~\cite{kim2023phase,jung2022aasist,liu2024multi} classifiers can acquire impressive results for \acrshort{fsd} task. The ResNet~\cite{he2016deep,xue2023learning} architecture, which has been successful in the image field~\cite{he2023interpretive,sun2022low}, has strong feature capture capabilities. Therefore, in~\cite{Zhang-et-al:scheme, Ling2021, zhang2021one}, authors propose a series of ResNet-based classifiers to detect fake speech. 
	To further enhance the generalization capability of the model, Gao et al.~\cite{Gaoetal} propose the Res2Net structure, which partitions channels into multiple groups and enables interaction through residual connections within each group to extract multi-scale features.
	Consequently, researchers~\cite{li2021channel, Xu-et-al:scheme} have made many beneficial attempts to use the Res2Net architecture for the \acrshort{fsd} task. 
	\begin{figure}[t]	
		\centering
		\includegraphics[width=\linewidth]{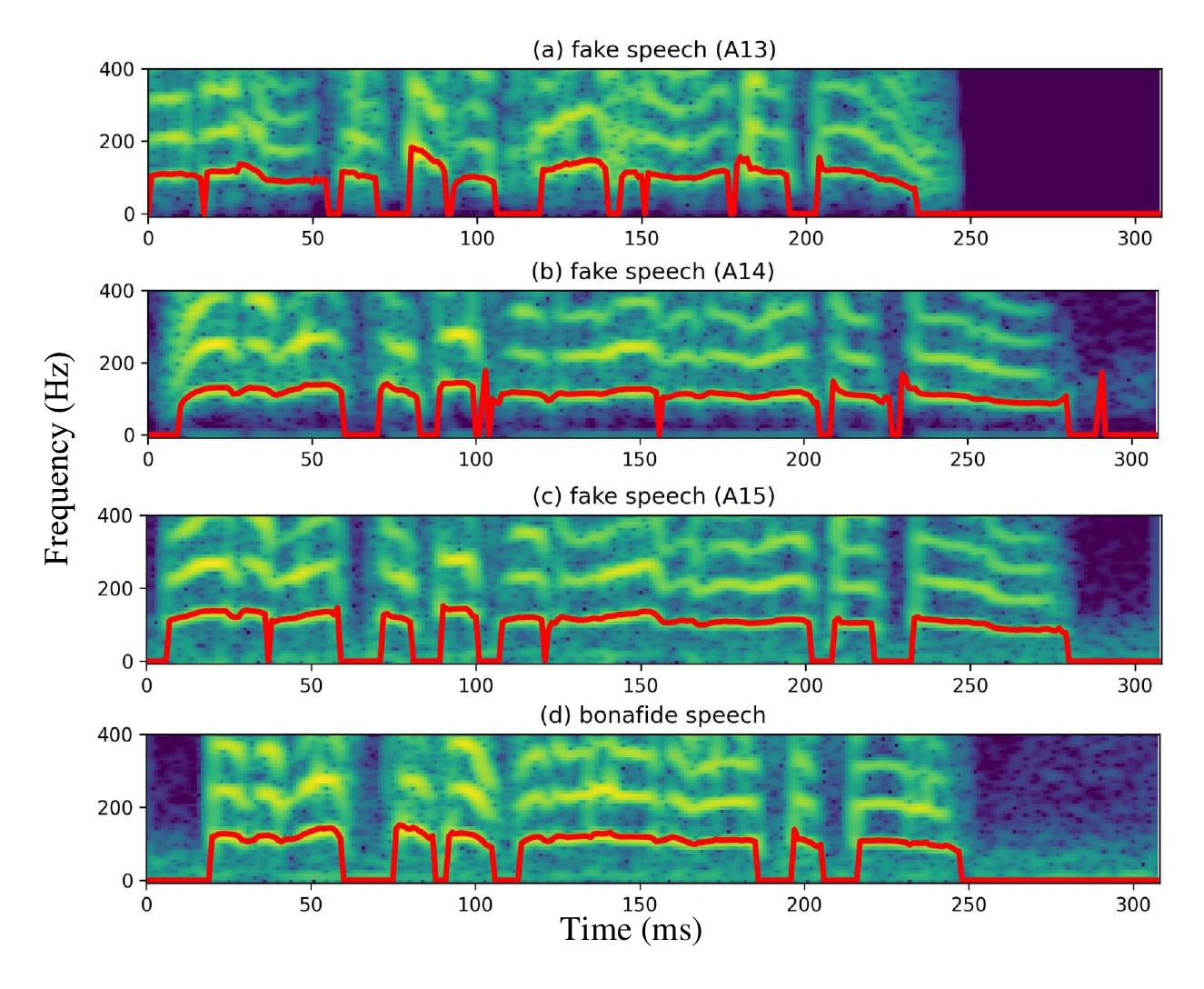}
		\caption {The spectrum and F0 distribution of three different types fake speech and the corresponding bonafide speech. Where the red line means the distribution of F0. 
				A13, A14 and A15 denote three different TTS algorithms drawn from the ASVspoof 2019 LA dataset, as described in~\cite{wang2020asvspoof}. The F0 distribution of these three different types of fake speech is distinctly different from the corresponding bonafide speech. This indicates that the F0 feature contains the discriminative information for the FSD task.}
		\label{fig:singleF0}
	\end{figure}
	However, the residual connections directly add information from the previous channel group to the next, and research~\cite{li2021channel} has shown that the cross-channel information can generate redundant information. Therefore, after aggregating information from multiple channel groups in Res2Net, the salient discriminative features may be interfered with by redundant information. These issues limit the performance of the FSD system.

Recent research in text-to-speech (TTS) has indicated that the fundamental frequency (F0) is very important for the quality of synthetic speech.
For example, in~\cite{lancucki2021fastpitch}, a new TTS model Fastpitch is proposed to predict the F0 contour during inference. Changed predictions make the generated speech more human-like. 
The field of Voice Conversion (VC) is also extensively exploring how to better model F0 when synthesizing speech. For example, in~\cite{qian2020f0}, the authors improve the autoencoder, which could better generate F0 contours consistent with the target speaker, as a way to significantly improve speech quality.
However, it is worth noting that the rhythm of bonafide speech is often difficult to replicate, which leads to the F0 of synthetic speech being very different from the F0 of real speech.
To compare the F0 distribution of fake speech and bonafide speech, Fig.~\ref{fig:singleF0} shows the spectrum and F0 distribution of three different fake voices and one bonafide speech, with the red line depicting the distribution of F0. 
From Fig.~\ref{fig:singleF0} we can find that the F0 distribution of these three different types of fake speech is distinctly different from the corresponding bonafide speech. This indicates that the F0 feature contains the discriminative information for the \acrshort{fsd} task.

	\begin{figure}[t]	
		\centering
		\includegraphics[width=\linewidth]{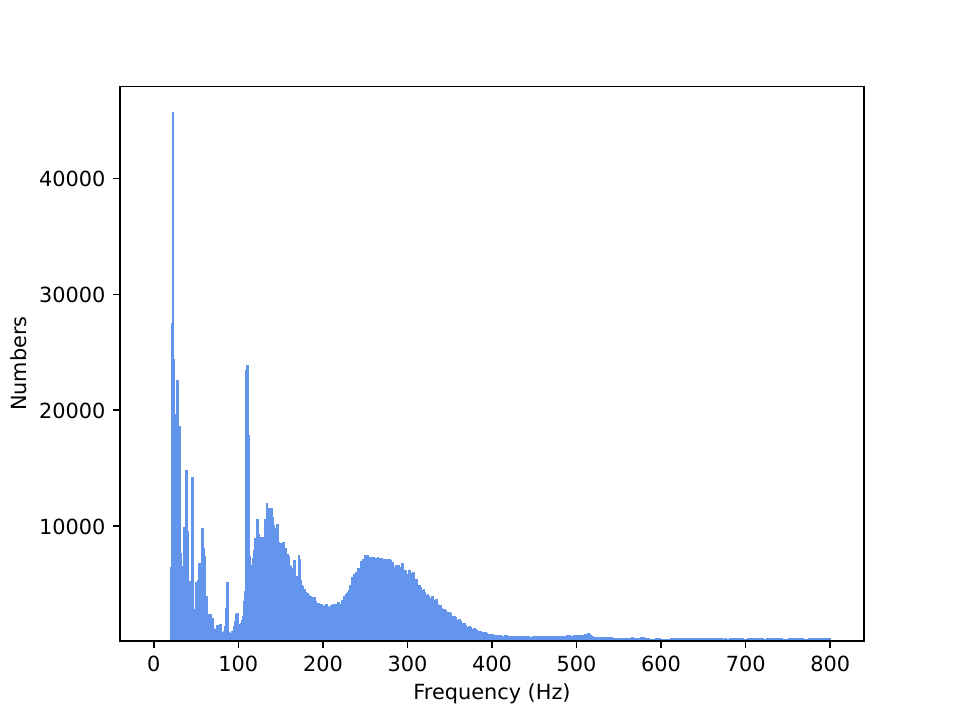}
		\caption {The frequency of F0 distribution in the ASVspoof 2019 LA training dataset. Where the abscissa is the frequency corresponding to F0, and the ordinate is the number of F0 at this frequency.}
		\label{fig:F0}
	\end{figure}
	
Unfortunately, F0 is difficult to model directly as a valid feature for FSD. To make full use of the discriminative information of F0, this paper proposes an F0 subband for \acrshort{fsd} task, which is the subband of amplitude spectrum. Fig.~\ref{fig:F0} shows the F0 distribution in the ASVspoof 2019 LA training dataset. 
From Fig.~\ref{fig:F0} we can find that most of the F0 is distributed between 0-400 Hz. Therefore, the frequency band containing most of the F0 is used as the F0 subband. 
Overall, compared to other acoustic features, the F0 subband contains a priori discriminative information, which avoids interference from redundant information.
In addition, to effectively model the F0 subband, we propose a novel spatial reconstructed local attention Res2Net (SR-LA Res2Net) for \acrshort{fsd}. Specifically, the Res2Net is used as the backbone network, which can capture the multi-scale information of the input feature. However, the gradual superposition of cross-channel group information will cause more artifacts to the spatial structure of the feature, and the redundant information generated by aggregation may obscure some important information. To address these problems, we design a spatial reconstructed (SR) block at the residual connection in Res2Net, which is used to reconstruct the spatial structure. Finally, the local attention (LA)\footnote{The module embedded in the Res2Net architecture proposed in this paper.} block is integrated at the bottom of Res2Net to focus on local information and capture the discriminative information of the F0 subband.

	The main contributions of this study can be summarized as two-fold. Firstly, we propose to use the F0 subband for the \acrshort{fsd} task, which is a very discriminative feature. Secondly, a novel SR-LA Res2Net architecture is designed to model the F0 subband, which can effectively solve the shortcomings of Res2Net when expanding feature receptive fields. The experimental results on the ASVspoof 2019 LA dataset show that our proposed method is very effective for the \acrshort{fsd} task, and it can acquire the state-of-the-art performance among all of the single systems. \textbf{The code\footnote{\textbf{https://github.com/JunXue-tech/SRLARes2NetF0Subband}} is available.}
	
	The rest of this article is arranged as follows. Section II introduces the related works. The proposed method is introduced in section III. Experiments and results are given in section IV. Section V shows the discussions. Section VI draws conclusions.

	\section{Related Works}
	
For the \acrshort{fsd} task, many studies~\cite{yang2019significance,zhang2021fake,chettri2020subband} have shown that different frequency bands have different effects. 
In~\cite{Zhang-et-al:scheme}, the authors focus on global channel attention using squeeze and extraction blocks and explore the impact of high frequency and low frequency subband for the \acrshort{fsd} task. The low frequency subband achieves good performance.
In~\cite{Ling2021}, the authors propose a frequency attention block and a channel attention block, which pay attention to the basic subband correlation and channel relationship, respectively.

In addition, many studies~\cite{lv2022fake,tak2021end,jung2022aasist,tak2021graph} are based on the Res2Net for \acrshort{fsd} and acquire quite good performances.
In~\cite{Xu-et-al:scheme}, the authors use Res2Net to enhance the system's generalization to unseen spoofing attacks and integrate squeeze and extract blocks to further improve performance.
In~\cite{li2021channel}, the authors propose a channel-wise gated Res2Net (CG-Res2Net), which dynamically adjusts the correlation between channels through a gating mechanism and suppresses channels with small correlations. It further enhances the generalization ability of the system against unseen spoofing attacks.

In this paper, we propose the F0 subband and SR-LA Res2Net for \acrshort{fsd}. Compared with Res2Net and CG-Res2Net, the proposed SR-LA Res2Net has better generalization ability.

	\begin{figure*}[!t]	
		\centering
		\includegraphics[width=1.0\textwidth]{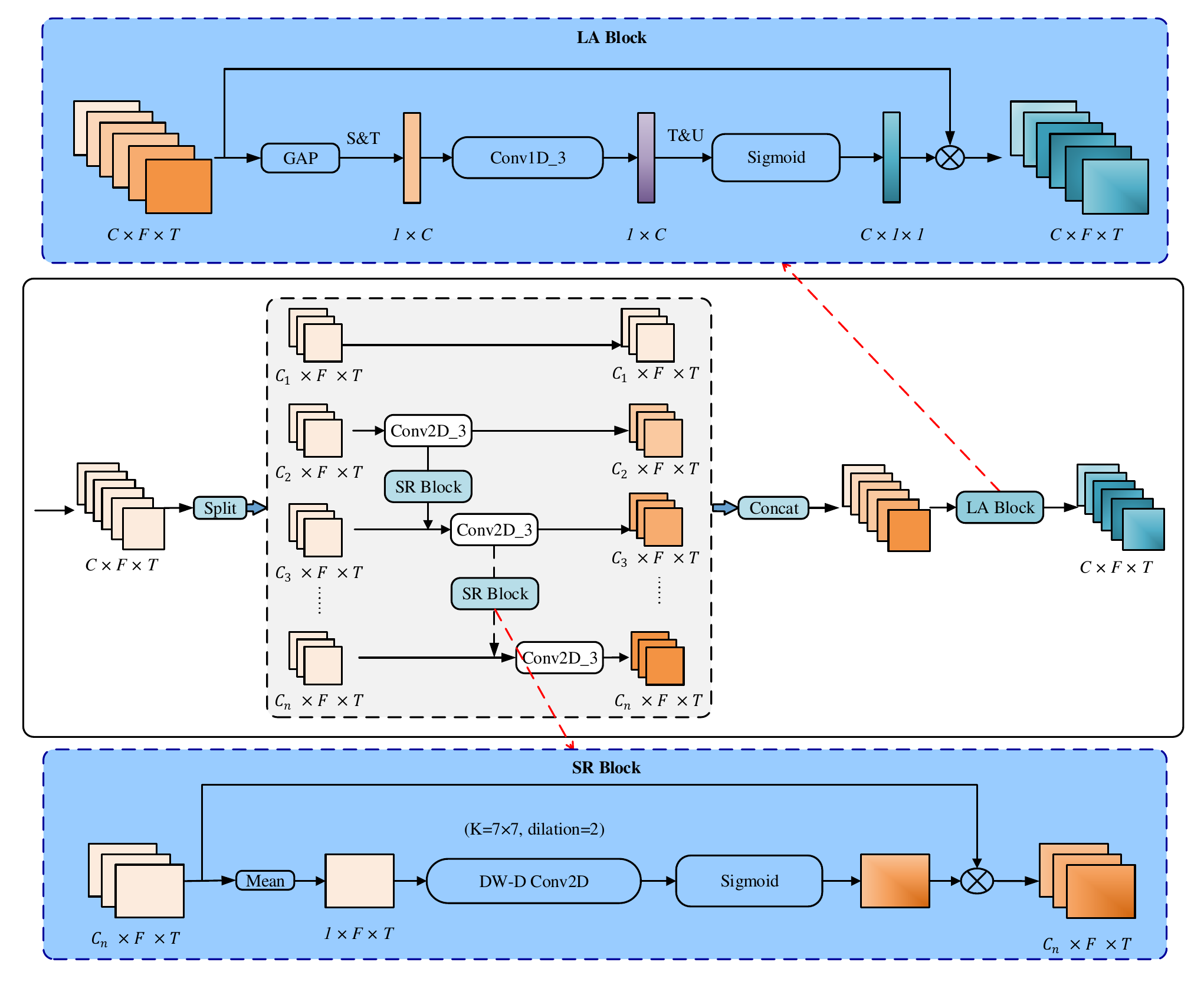}
		\caption {The schematic diagram of the proposed SR-LA Res2Net architecture. The spatial reconstructed (SR) block is used to remove the spatial structure artifacts of the channel group, and the local attention (LA) block aims to highlight important information and remove redundant information. }
		\label{fig:Res2Net}
	\end{figure*}

	\section{The Proposed F0 Subband with SR-LA Res2Net}
	
	In this paper, we propose an F0 subband with SR-LA Res2Net for \acrshort{fsd}. Because the F0 of synthetic speech is very different from the real one. Therefore, we think the F0 contains the discriminative information and apply the F0 subband as the input feature for \acrshort{fsd}. To further improve the performance of \acrshort{fsd}, we propose the SR-LA Res2Net to model the F0 subband feature, which can effectively solve the shortcomings of Res2Net when expanding feature receptive fields.

	\subsection{F0 Subband}
	
	To make full use of the discriminative information of F0, we extract the F0 subband based on the LPS. Specifically, the short-time Fourier transform (STFT) is used to convert the time domain raw waveform \(\textbf{x}[k]\) into the time-frequency (T-F) domain.
	\begin{equation}
		\boldsymbol{X}_\text{r}[t,f] + j\cdot\boldsymbol{X}_\text{i}[t,f] = STFT(\textbf{x}[k])
		\label{eq1}
	\end{equation}
	where \(k\) is the time index of raw waveform \(\textbf{x}[k]\). \(STFT\) means the operation of STFT. \(\boldsymbol{X}_\text{r} \in{\mathbb{R}^{{F}\times{T}}}\) and \(\boldsymbol{X}_\text{i} \in{\mathbb{R}^{{F}\times{T}}}\) are the corresponding real and imaginary part of STFT, respectively. \(t\) is the index of time frame and \(f\) is the index of frequency bin. \(F\) and \(T\) are the number of frequency bins and time frames, respectively. For convenience, \((t,f)\) is omitted from the following formulas in this paper.
	
	The full frequency bands of \(\boldsymbol{LPS}_\text{full}\) can be acquired as follows:
	\begin{equation}
		\boldsymbol{LPS}_{\text{full}} = \log\sqrt{(\boldsymbol{X}_\text{r})^2 + (\boldsymbol{X}_\text{i})^2} \in{\mathbb{R}^{{F}\times{T}}}
		\label{eq2}
	\end{equation}

	From Fig.~\ref{fig:F0} we can find that most of the F0 is distributed between 0-400 Hz. Therefore, the 0-400 Hz of LPS is applied as our F0 subband \(\boldsymbol{LPS}_\text{F0}\). 
	\begin{equation}
		\boldsymbol{LPS}_\text{F0}=\boldsymbol{LPS}_{\text {0-400 Hz}}
		\label{eq3}
	\end{equation}

	\subsection{Model Architecture}
	
To effectively model the F0 subband and improve the performance of \acrshort{fsd}, we propose the SR-LA Res2Net architecture. 
Fig.~\ref{fig:Res2Net} shows the schematic diagram of the proposed SR-LA Res2Net architecture. Firstly, to extract the multi-scale information of the F0 subband, the Res2Net is used as the backbone. However, when the channel group is constantly superimposed, the Res2Net may generate redundant information so that much important information may be lost. To address this issue, the SR block is proposed at the residual connection between channel groups, which can restore the spatial structure. In addition, an LA block is designed at the bottom of Res2Net to pay attention to local information and remove the influence of redundant information. Therefore, our proposed SR-LA Res2Net can further remove spatial artifacts and redundant information while extracting multi-scale features, thereby improving the generalization ability of the model to unseen spoofing attacks.
	
	\subsubsection{\(\textbf {The Res2Net Architecture}\)} 
	
	The ResNet has been applied in various fields as soon as it was proposed and has achieved great performance. 
	Even if ResNet's residual connections can reduce the impact of network depth, just increasing the network depth does not improve the performance of the model very well. So Gao et al.~\cite{Gaoetal} proposed the Res2Net architecture, which obtains multi-scale features through the information transfer of channel groups.
	Firstly, to expand the range of interaction between channel groups, the input features are divided into $n$ subsets according to the channel dimension after \(1\times1\) convolution, denoted as $s_{i}$, where $i \in\{1,2, \ldots, n\}$. As for $s_{1}$, it does not undergo any processing. As for $s_{2}$, it is directly output after a \(3\times3\) convolution $K_{2}(\cdot)$. As for $s_{3}$ to $s_{n}$, each $s_{i}$ needs to be added to the output of $K_{i-1}$ before passing through $K_{i}(\cdot)$. 
	This process can be formulated as follows:
	\begin{equation}
		y_{i}= \begin{cases}s_{i}, & i=1 \\ K_{i}\left(s_{i}\right), & i=2 \\ K_{i}\left(s_{i}+y_{i-1}\right), & 3 \leq i \leq n \end{cases}
	\end{equation}
	where $n$ is defined as the scale dimension, indicating the number of channel groups applied to split feature maps. 
	
	Therefore, the Res2Net can increase the interaction between channel groups through residual connections in the block. Through the residual connections in the block, each channel group obtains a different amount of information, thereby it can generate multiple-scale features. Such a multi-scale representation increases the receptive field. Finally, all channel groups are aggregated and the original channel size is maintained by the \(1\times1\) convolution.

	\subsubsection{\(\textbf {Spatial Reconstructed Block}\)}
	
	The Res2Net has a strong feature representation ability, which relies on the information transfer of multiple internal channel groups. However, as the feature information continues to be superimposed, more artifacts will appear in its spatial structure. This greatly affects the performance of the Res2Net model. Inspired by~\cite{woo2018cbam}, we design a SR block for residual connection between channel groups. It aims to reconstruct the feature space and remove its artifacts when the information is passed to the next channel group.
	
	Firstly, to reduce subsequent parameter computations, we compress the channel dimension:
	\begin{equation}
		\mathcal{F}_{\mathrm{Mean}} \in \mathbb{R}^{1 \times F \times T}=\operatorname{Mean}\left(\mathcal{F}_{\mathrm{in}} \in \mathbb{R}^{C \times F \times T}\right)
	\end{equation}
	where $\operatorname{Mean}(\cdot)$ means the mean operation, $\mathcal{F}_\text{in}$ is the input feature, and \(C\) is the number of channels.
	
	Then, to further expand the receptive field, the depth-wise dilation convolution is applied:
	\begin{equation}
		\mathcal{F}_\text{DW-DC} \in \mathbb{R}^{1 \times F \times T}=\operatorname{DW-DC}\left(\mathcal{F}_{\text {Mean }} \in \mathbb{R}^{1 \times F \times T}\right)
	\end{equation}
	
	where $\operatorname{DW-DC}$ represents the operation of depth-wise dilation convolution.

	Finally, $	\mathcal{F}_\text{DW-DC} \in \mathbb{R}^{1 \times F \times T}$ reconstructs the feature space through the sigmoid layer. Then it multiplies with the input feature $\mathcal{F}_\text{in}$:
	\begin{equation}
		\mathcal{F}_\text{sr} \in \mathbb{R}^{C \times F \times T} =\mathcal{F}_{\mathrm{in}} \otimes \operatorname{sigmoid}\left(\mathcal{F}_\text{DW-DC} \in \mathbb{R}^{\mathrm{1} \times F \times T}\right)
	\end{equation}
	where $\mathcal{F}_\text{sr}$ represents the reconstructed feature space. \(\otimes\) denotes the element-wise multiplication.

	\subsubsection{\(\textbf {Local Attention Block}\)}
	
	Although Res2Net can obtain multi-scale global information, due to the uneven interaction information between the channel groups, this uneven global interaction may lead to information redundancy and the useful information may not be focused. To address this problem, motivated by~\cite{Wang-et-al:scheme}, the LA block is applied to focus on the local information.
	
	Firstly, the global average pooling (GAP) is used to squeeze the dimensions of the input features:
	\begin{equation}
		\mathcal{F}_{\text{GAP}} \in \mathbb{R}^{C \times 1 \times 1}=\operatorname{G A P}\left(\mathcal{F}_{\text{g}} \in \mathbb{R}^{C \times F \times T}\right)
	\end{equation}
	Where $\operatorname{GAP}(\cdot)$ represents a GAP operation, $\mathcal{F}_{g}$  is the output of Res2Net feature aggregation.
	
	In order to squeeze the one-dimensional channel of the convolution, squeeze and transpose operations are then applied:
	\begin{equation}
		\mathcal{F}_\text{S\&T} \in \mathbb{R}^{1 \times C}=\operatorname{S\&T} \left(\mathcal{F}_\text{GAP} \in \mathbb{R}^{C \times 1 \times 1}\right)
	\end{equation}
	where $\operatorname{S\&T(\cdot)}$ means the operation of squeeze and transpose.
	
	
	\begin{table}[t]
		\caption{The detailed information of ASVspoof 2019 LA Dataset. Where the ``utt." means the number of utterance.}
		\label{tab:LAset}
		\renewcommand\arraystretch{1.2}
		\begin{tabular}{p{30mm}ccc}
			\hline \toprule\multirow{2}{*}{Partition } & Bonafide & Spoof & Spoof \\
			\cline { 2 - 4 } & utt. & utt. & attacks type. \\
			\hline Train. & 2580 & 22800 & A01-A06 \\
			\hline Dev. & 2548 & 22296 & A01-A06 \\
			\hline Eval. & 7355 & 63882 & A07-A19 \\
			\hline \toprule
		\end{tabular}
	\end{table}
	
In~\cite{hu2019squeeze}, the authors proposed the Squeeze-and-Excitation (SE) block, which is different from the LA block in that they use two fully connected layers to learn global channel attention. The first FC layer is used for dimensionality reduction, and the second FC layer is used to restore the dimensionality.
Although the parameters of this method are reduced by the dimensionality reduction, the complexity of the model is still very high. In addition, the dimensionality reduction can affect the performance of the model. To address this issue, motivated by~\cite{Wang-et-al:scheme}, we apply the one-dimensional convolution to acquire the local attention information. The details are as follows:
	\begin{equation}
		\mathcal{F}_\text{Conv}=\operatorname{Conv}\left(\mathcal{F}_\text{S\&T} \in \mathbb{R}^{1 \times C}\right)
	\end{equation}
	where the \(\operatorname{Conv}(\cdot)\) is the operation of one-dimensional convolution. 
	
	Then, the feature size is gradually restored by the transpose and unsqueeze operations, which is defined as follows:
	\begin{equation}
		\mathcal{F}_\text{T\&U} \in \mathbb{R}^{C \times 1 \times 1}= \operatorname{T\&U}\left(\mathcal{F}_\text{Conv} \in \mathbb{R}^{1 \times \mathrm{C}}\right)
	\end{equation}
	where $\operatorname{T\&U(\cdot)}$ means the operation of transpose and unsqueeze.
	
	Finally, the $\mathcal{F}_\text{T\&U} \in \mathbb{R}^{\mathrm{C} \times 1 \times 1}$ is passed by the sigmoid layer to acquire the vector of attention weight. The finally local attention vector can be obtained by multiplying the attention weight and the input feature $\mathcal{F}_\text{g}$ of local attention block.
	\begin{equation}
		\mathcal{F}_\text{la}=\mathcal{F}_\text{g} \otimes \operatorname{sigmoid}\left(\mathcal{F}_\text{T\&U} \in \mathbb{R}^{\mathrm{C} \times 1 \times 1}\right)
	\end{equation}
	where the $\mathcal{F}_\text{la}$ denotes the output of local attention block.

	\begin{table}[t]
		\caption{The proposed SR-LA Res2Net model architecture and configuration. Dimensions refer to (channels, frequency, and time). Batch normalization (BN) and Rectified Linear Unit (ReLU), SR and LA are the spatial reconstruction block and the local attention block, respectively.}
		\label{tab:SR-LA Res2Net}
		\setlength{\tabcolsep}{0.96mm}
		\begin{tabular}{ccc}
			\hline
			Layer                            & Input: 27000 samples & Output shape  \\ \hline
			Front-end                        & F0 subband           & (45,600)(F,T) \\ \hline
			\multirow{3}{*}{Post-processing} & Add channel          & (1,45,600)    \\
			& Conv2D\_1            & (16,45,600)   \\
			& BN \& ReLU           &               \\ \hline
			Res2-block                       &  $\begin{matrix}
				2 \times\left\{\begin{array}{c}
					
					\text { Conv2D\_1 } \\
					\text { Conv2D\_3 \& SR}  \\
					
					\text { Conv2D\_1 }  \\
					\text {LA}

				\end{array}\right\}		 
			\end{matrix}$            & (32,45,600)   \\ \hline
			Res2-block                       &  $\begin{matrix}
				2 \times\left\{\begin{array}{c}
					
					\text { Conv2D\_1 } \\
					\text { Conv2D\_3 \& SR}  \\
					
					\text { Conv2D\_1 }  \\
					\text {LA}

				\end{array}\right\}		 
			\end{matrix}$            & (64,23,300)   \\ \hline
			Res2-block                       &  $\begin{matrix}
				2 \times\left\{\begin{array}{c}
					
					\text {Conv2D\_1}\\
					\text { Conv2D\_3 \& SR}  \\
					
					\text {Conv2D\_1 }  \\
					\text {LA}

				\end{array}\right\}		 
			\end{matrix}$            & (128,12,150)  \\ \hline
			Res2-block                       &  $\begin{matrix}
				2 \times\left\{\begin{array}{c}
					
					\text { Conv2D\_1} \\
					\text { Conv2D\_3 \& SR}  \\
					
					\text { Conv2D\_1 }  \\
					\text {LA}

				\end{array}\right\}		 
			\end{matrix}$            & (256,6,75)    \\ \hline
			\multirow{2}{*}{Output}          & Avgpool2D(1,1)       & (256,1,1)     \\
			& AngleLinear          & 2             \\ \hline
		\end{tabular}
	\end{table}

	\section{Experiments and Results}
	
	\subsection{Dataset}
	
	We conduct our experiments on the ASVspoof 2019 LA database and the ASVspoof 2021 LA database.
	
	\subsubsection{ASVspoof 2019 LA Database}
	
	ASVspoof 2019 LA\footnote{https://datashare.ed.ac.uk/handle/10283/3336} mainly has 19 spoofing attack algorithms (A01-A19), including three types of spoofing attacks: TTS, voice conversion (VC), and audio playback. Table~\ref{tab:LAset} details the components of the ASVspoof 2019 LA dataset. It can be seen that the LA subset has three parts: training, development, and evaluation.
	Among them, the training set is used to train the model, the development set is used to select the best performing model in training, and finally, the model performance is evaluated through the evaluation set. The training set and development set mainly include four TTS and two VC algorithms, namely A01-A06. To better evaluate the performance of the system, unseen spoofing attacks were added to the evaluation set, including two known spoofing attacks (A16 and A19) and 11 unseen spoofing attacks (A07-A15, A17, and A18).

			\begin{figure}[!t]	
		\centering
		/	\includegraphics[width=0.43\textwidth]{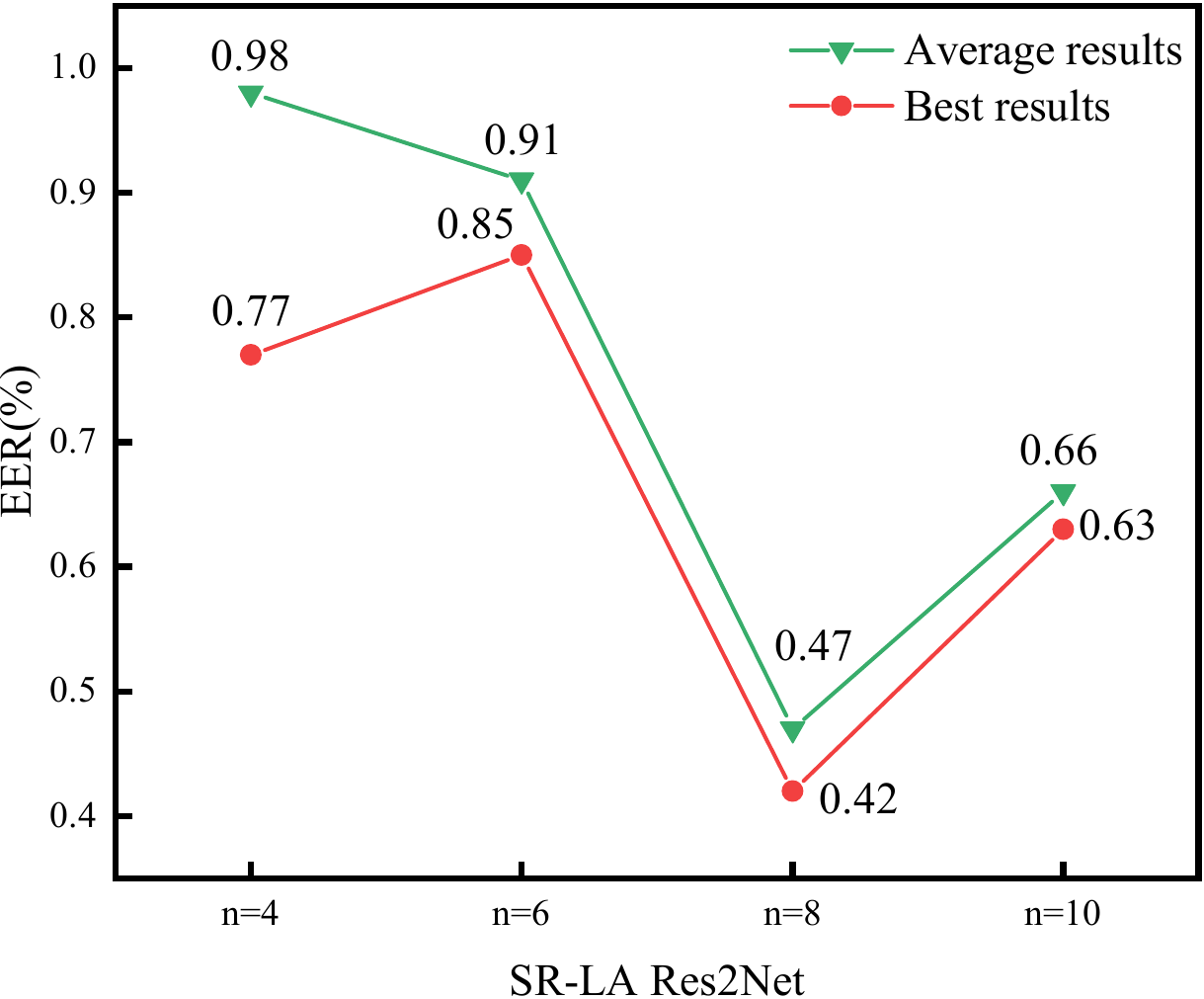}
		\caption {The EER results of SR-LA Res2Net for different number of channel groups (n). To avoid randomness in the experiments, the results are averaged for the three runs. Where the green line meas the average results and the red line denotes the the best results of the three runs.}
		\label{fig:n_value}
	\end{figure}
	
	\subsubsection{ASVspoof 2021 LA Database}
	
The difference between ASVspoof 2019 and 2021\footnote{https://zenodo.org/records/4837263} LA database is the evaluation set. Therefore, the ASVspoof 2019 LA training and dev sets are used to train the model. The evaluation set contains about 180,000 utterances transmitted through real telephone systems with different bandwidths and different codecs. The transmission interference of this data set greatly affects the performance of the \acrshort{fsd} system and makes it more challenging.

To quantitatively evaluate the performance of different \acrshort{fsd} systems, EER and the minimum normalized tandem detection cost function (min t-DCF) are applied. EER is the working point where the false rejection rate (FRR) and false acceptance rate (FAR) are equal.

	\subsection{Experimental Setup}

First, we perform STFT operations on the original audio waveform. We use Blackman as the window function of the STFT and set the window length and hop length to 1728 and 130, respectively, to obtain a spectrogram of size 865. Then, we fix the number of frames to 600 by truncating and concatenating. So the feature dimension is 865×600.
We take the 0-400 Hz LPS feature as the F0 subband, so the corresponding frequency dimension is 45. Therefore, the first 0-45 dimensions are taken as the F0 subband features, and 45×600 is obtained by cutting the above features.

		\textit{Network architecture:}
		In this paper, Res2Net~\footnote{https://github.com/Res2Net} is used as the backbone network, the proposed SR block is embedded in the internal channel residual connections of Res2Net, and the proposed LA block is embedded after the channel aggregation of Res2Net. As shown in Figs~\ref{fig:Res2Net} and Table~\ref{tab:SR-LA Res2Net}, the details including convolution kernel, channels, and repetition counts are provided.
		In addition, we use Adam as the optimizer, and the parameters of the optimizer are set to:  $\beta_{1}=0.9$, $\beta_{2}=0.98$, $\epsilon=10^{-9}$ and weight decay is $10^{-4}$. The number of the epoch is 32.
		Fig.~\ref{fig:n_value} shows the EER results for different numbers of channel groups (n) based on the SR-LA Res2Net architecture. From Fig.~\ref{fig:n_value}, we can see that the best performance is achieved for n=8, so we set the number of channel groups to 8 in the experiment.

		In addition, since the ASVspoof 2021 LA dataset was interfered with by transmissions such as telephone communications, we used the Rawboost\footnote{https://github.com/TakHemlata/RawBoost-antispoofing}~\cite{tak2022rawboost} data enhancement method when training the evaluation used for the ASVspoof 2021 LA dataset. Specifically, we added impulse signal-independent additive noise and stationary signal-independent additive noise to the original waveform.

	\begin{figure}[!t]	
		\centering
		\includegraphics[width=0.45\textwidth]{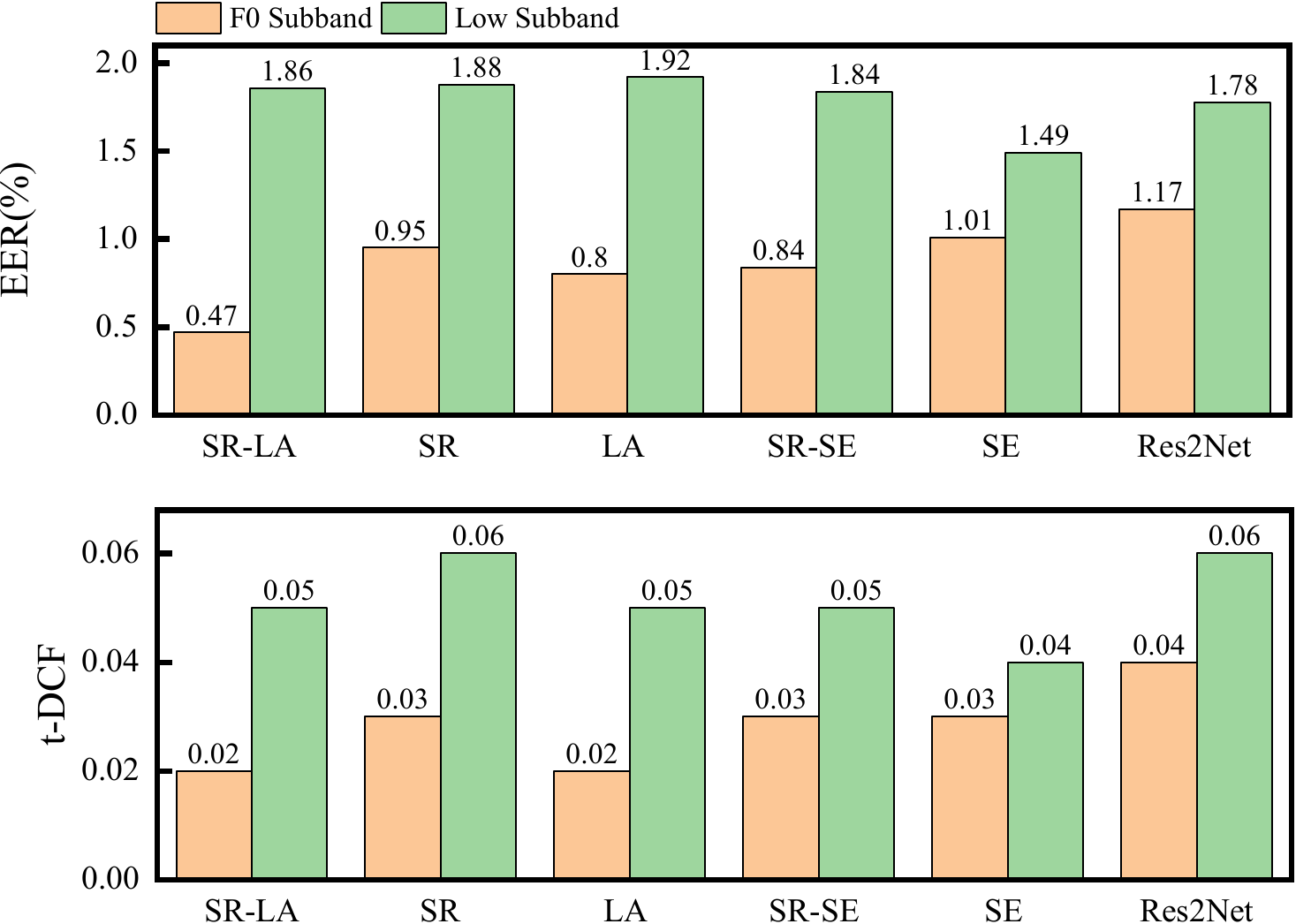}
		
		\caption {The EER and t-DCF results for our proposed different systems in the ASVspoof 2019 LA evaluation set (A07-A19). Where SR, LA, and SE are the different components embedded in the Res2Net.}
		\label{fig:F0_L}
	\end{figure}

	\subsection{Experimental Results on ASVspoof 2019 LA dataset}
	\subsubsection{Effectiveness of the F0 Subband}\
	
\begin{figure*}[!t]	
		\centering
		\includegraphics[width=1.0\textwidth]{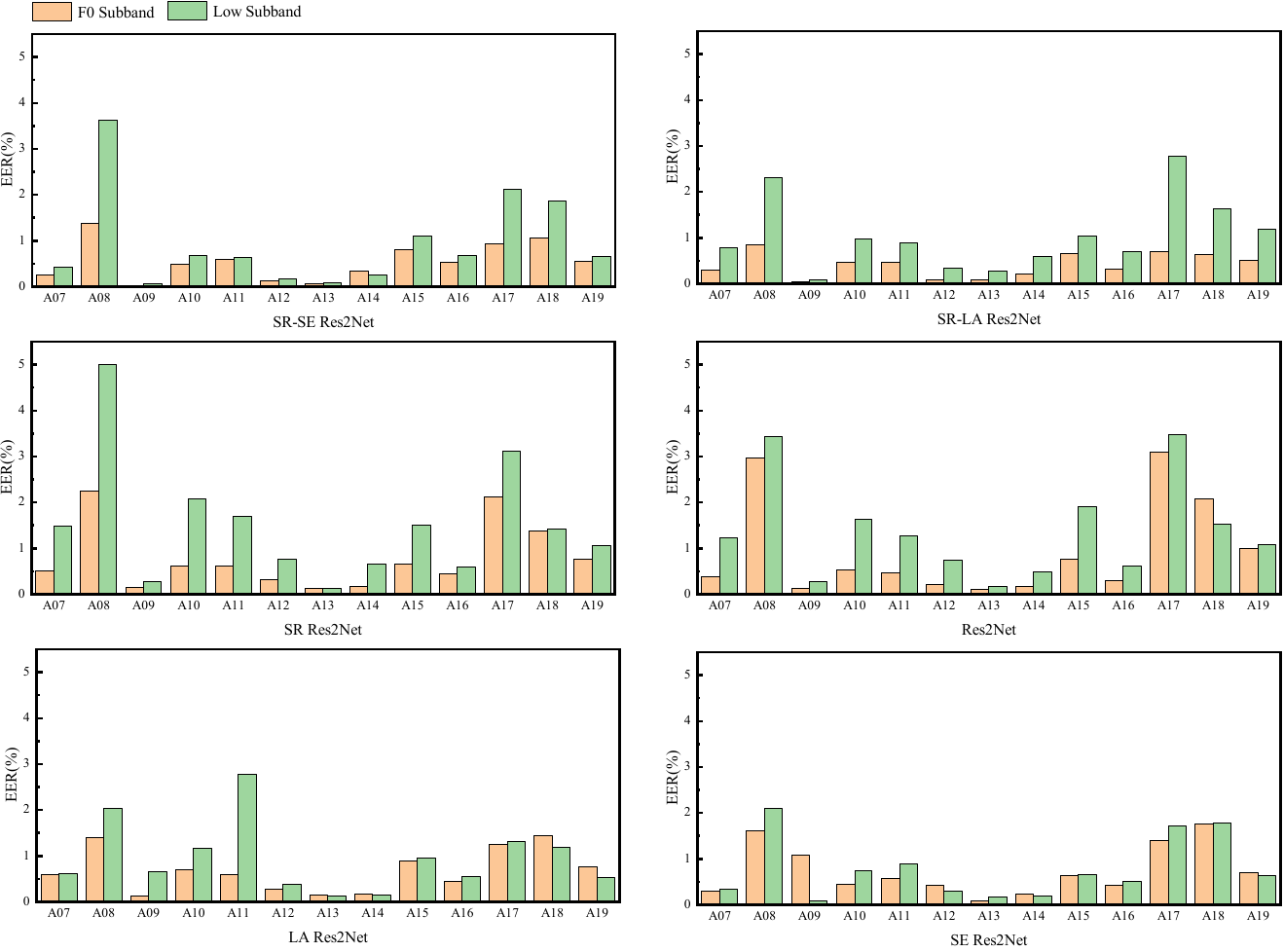}
		\caption {EER results of our proposed different systems on the ASVspoof 2019 LA evaluation set (A07-A19). The EER of each spoof method is calculated separately.}
		\label{fig:A07}
	\end{figure*}

	This section evaluates the effectiveness of F0 subband feature on different network structures. Fig~\ref{fig:F0_L} shows the minimal t-DCF and EER results for the different systems we proposed. To avoid randomness in the experiments, the results are averaged for the three runs, with the best of the three runs in parentheses.
	``F0'' represents based on the F0 subband feature, and ``L'' represents based on the low frequency (0-4000 Hz)~\cite{Zhang-et-al:scheme} subband feature. The first six lines are the results of the F0 subband feature. The last six lines are experimental results based on low frequency subband feature.
	Fig~\ref{fig:A07} shows the EER histograms based on the F0 subband and low subband features, with EERs calculated separately for different attack types.
	
	From the Fig~\ref{fig:F0_L} and Fig.\ref{fig:A07}, we can see the following.\
	
	\begin{table}[t]
		\caption{Results of our proposed ablation experiments for different components. The results are the average of three runs, with the best of the three results in parentheses.}
		\label{tab:result2}
		\setlength{\tabcolsep}{0.96mm}
		\renewcommand\arraystretch{1.2}
		\centering
		\begin{tabular}{p{40mm}cc}
			\hline \toprule \centering Systems & t-DCF & EER(\%) \\ \hline
			\centering SR-LA Res2Net (F0) & $\textbf {0.0159(0.0143)}$ & $\textbf {0.47(0.42)}$  \\ 
			\centering SR-SE Res2Net (F0)& $0.0270(0.0227)$ & $0.84(0.74) $ \\
			\centering SR Res2Net (F0) &  $ 0.0306(0.0302)$ & $0.96(0.95)$ \\
			\centering LA Res2Net (F0) &  $0.0246(0.0229)$ & $0.80(0.77)$  \\
			\centering SE Res2Net (F0) & $0.0310(0.0292)$ & $1.01(0.95) $ \\
			\centering Res2Net (F0) & $0.0353(0.0335)$ & $1.17(1.14)$ \\
			\centering LA ResNet (F0) & $0.0388(0.0364)$ & $1.26(1.14)$ \\
			\centering SE ResNet (F0)& $0.0424(0.0392)$ & $1.36(1.23)$ \\
			\centering ResNet (F0)& $0.0493(0.0406)$ & $1.64(1.34) $ \\
			\hline \toprule
		\end{tabular}
	\end{table}

	(1) In the LPS features, the performance of the F0 subband is better than that of the lower subband features in all cases. For example, in our proposed state-of-the-art classifier-SR-LA Res2Net, the average EER of its F0 subband is 0.47\%, while the EER of the low subband is 1.86\%. Even though the low subband (0-4000Hz) has ten times more band information than F0 (0-400Hz), it performs much worse in the \acrshort{fsd} task. This is because the main discriminative information is concentrated in the F0 sub-band, and the other frequency band information may make it overfitting. The experimental results show that the F0 subband is an important identification feature.
	
	(2) The F0 subband feature has general applicability for different types of attacks. For low subband features, the attack types of A08 (neural waveform), A17 (waveform filtering), and A18 (vocoder) are difficult to detect. 
	For example, ~\cite{al2018continuous} proposed a source-filter based vocoder, in order to refine the output of the contF0 estimation, the authors used post-processing to reduce the unwanted vocalized components of unvoiced speech, resulting in a smoother contF0 trace.
	It can be seen in~Fig.\ref{fig:A07} that the F0 subband has good performance for different attack types, even for the notorious attack type like A17. Overall, the F0 subband features outperform the lower subbands on different classifiers.
	
	(3) SR-LA Res2Net classifier can fully exploit the discriminatory ability of F0 subbands. For example, from ResNet to SR Res2Net and then to SR-LA Res2Net, attacks such as A08 and A17 are greatly optimized under the F0 subband, but it is more difficult to develop for the low subband, which may be due to the interference of having a lot of redundant information in the low subband.

		\begin{figure}[!t]	
		\centering
		\includegraphics[width=0.46\textwidth]{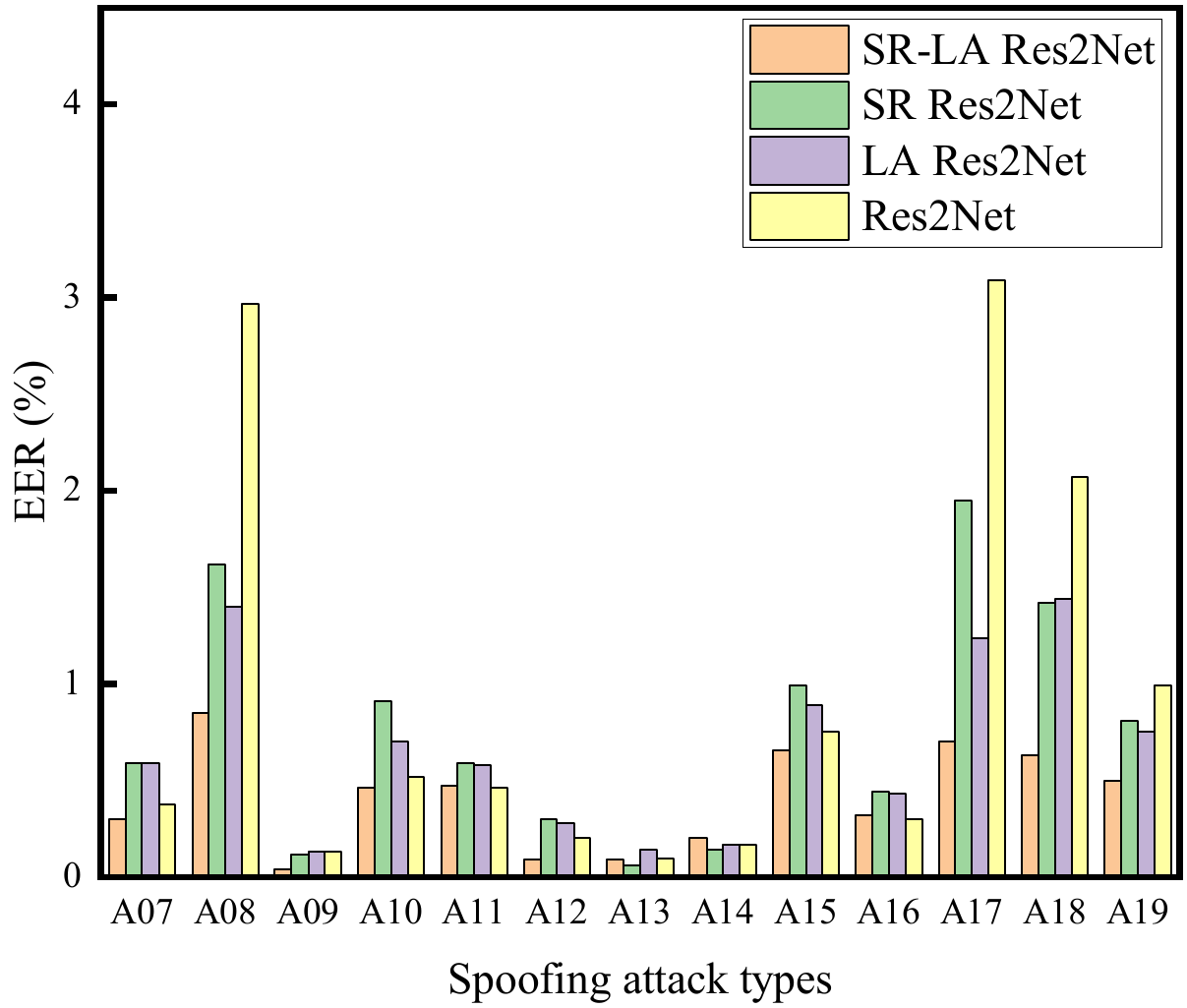}
		\caption {EER(\%) results for the evaluation subset (A07-A19). The EER(\%) of each spoof method is calculated separately.}
		\label{fig:abolation}
	\end{figure}

	\vspace{1ex}
	\subsubsection{Effectiveness of the SR-LA Res2Net Architecture}\

	To verify the effectiveness of our proposed \acrshort{fsd} system based on the F0 subband and the SR-LA Res2Net, we performed a series of ablation experiments. Table~\ref{tab:result2} shows the min t-DCF and EER results of our proposed different systems.
	In addition, to validate the effectiveness of our proposed SR-LA Res2Net, we used some recently published advanced network to model the F0 subband. Fig.~\ref{fig:abolation} shows the EER for the different networks.

	Firstly, the multi-scale feature representation of F0 subband is an effective way to improve the performance of the pseudo-speech detection system. the EER result of "ResNet (F0)" is 1.64\%, while the EER result of "Res2Net (F0)" is 1.17\%. This is due to the fact that Res2Net is designed with residual connections within the channel group so that the model can learn information at different scales to discriminate.
	However, when the Res2Net extracts multi-scale features, the larger the number of channel groups, the more information is superimposed, and the spatial structure of the features will also have more artifacts, which affects the ability of the model to capture fundamental discriminative information. Therefore, we design the spatial reconstructed block to be integrated into the residual connections of Res2Net to reconstruct the feature space before transferring the information. Experimental results show that SR-Res2Net (F0) has a good performance improvement over Res2Net (F0). This suggests that the spatial reconstruction mechanism helps remove spatial structure artifacts and further improves the performance of the model.


	\begin{figure}[!t]	
		\centering
		\includegraphics[width=0.46\textwidth]{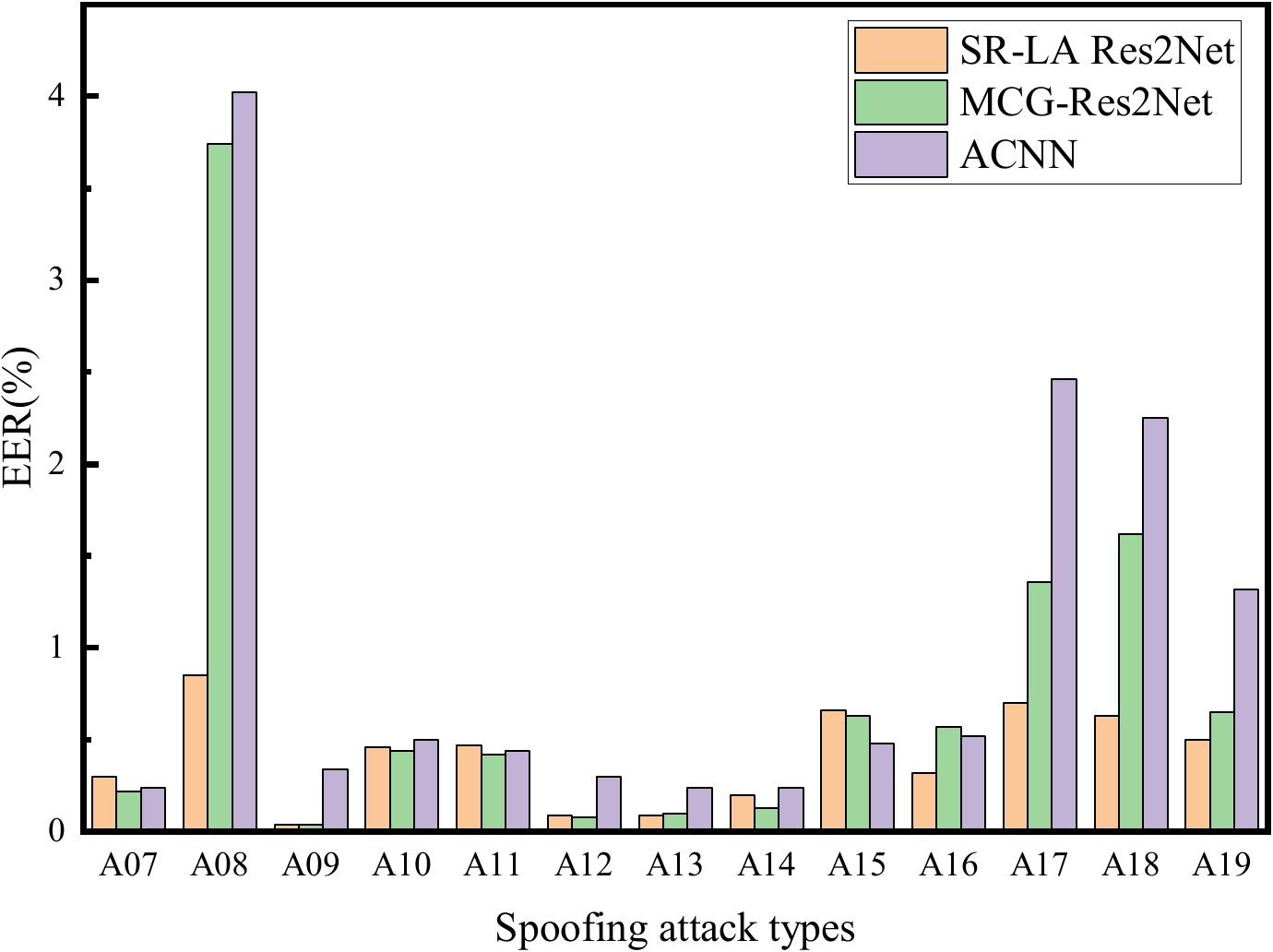}
		\caption {EER results of the F0 subband feature on other advanced classifiers.}
		\label{fig:Other}
	\end{figure}
	
	\begin{table*}[t]	
			\caption{Comparison of the results of F0 subband features on other advanced classifiers.}
		\label{tab:result3_parm}
		\renewcommand\arraystretch{1.2}
		\centering

		\begin{tabular}{ccccc}
			
			\hline
			Systems                      & Input      & t-DCF  & EER(\%)    & \#Parm                 \\ \hline
			\multirow{2}{*}{ACNN~\cite{Ling2021}}        & FFT        & 0.0510 & 1.87   & \multirow{2}{*}{1.04M} \\ \cline{2-4}
			& F0 subband & \textbf{0.0454} & \textbf{1.46}   &                        \\ \hline

			\multirow{2}{*}{MCG-Res2Net~\cite{li2021channel}} & CQT       & 0.0520 & 1.78   & \multirow{2}{*}{1.09M} \\ \cline{2-4}
			& F0 subband & \textbf{0.0299} & \textbf{1.03}   &   \\ \hline
			
			\multirow{2}{*}{LCNN~\cite{Lavrentyeva-et-al:scheme}} & FFT       & 0.1028 & 4.53   & \multirow{2}{*}{0.78M} \\ \cline{2-4}
& F0 subband & \textbf{0.0417} & \textbf{1.30}   &   \\ \hline

			\multirow{2}{*}{ResNet18-L-FM~\cite{chen2020generalization}} & LFBs       & 0.0520 & 1.81   & \multirow{2}{*}{0.68M} \\ \cline{2-4}
& F0 subband & \textbf{0.0465} & \textbf{1.48}   &   \\ \hline
			                     
			SR-LA Res2Net ( \textbf{Ours})          & F0 subband &  \textbf{0.0159}  &\textbf{0.47}  & 0.95M                  \\ \hline 
		\end{tabular}
	\end{table*}	
	
		\begin{figure}[t]	
		\centering
		\includegraphics[width=0.8\linewidth]{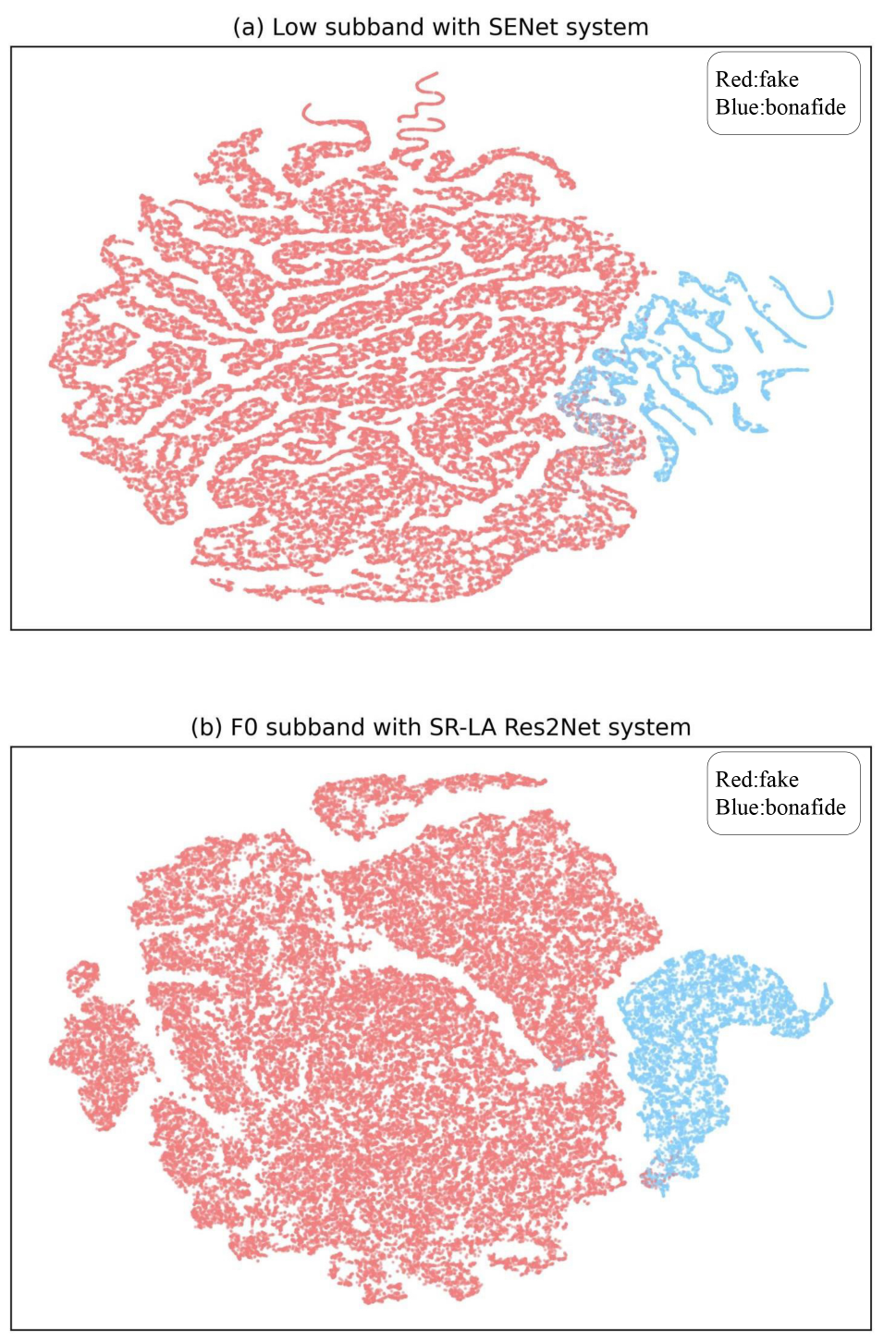}
		\caption {Subplot (a) shows the t-SNE visualization for the low-frequency subband and SENet system, and subplot (b) shows the t-SNE visualization for the F0 subband and SR-LA Res2Net system. The blue dots represent real speech and the red dots represent false speech.}
		\label{fig:t-sne}
	\end{figure}
	
	Next, we integrate a local attention block at the bottom of Res2Net. This is because the deepest channel group of the Res2Net superimposes all the information, and the rest also superimposes a lot of information, which leads to the generation of a lot of redundant information and further covers the important discriminative information. We propose to restore the weights of important information through local attention blocks after this information is aggregated. Specifically, we also compare the performance of local attention (LA) and global attention (SE)~\cite{hu2019squeeze}, and the experimental results show that local attention is better than global attention. We think there are two reasons: \ding{172} When generating the attention map, much long-distance information in the global information cannot accurately capture its specific connection, and the short-distance information can better judge the weight of the central information; \ding{173} The global attention block uses two fully connected layers, which are used for dimensionality reduction and expansion, respectively. The dimensionality reduction operation may result in the loss of some information, which can affect the discriminability of the model. According to the experimental results of integrating local attention blocks and global attention blocks respectively, LA Res2Net achieves an EER result of 0.80\%, and SR-LA Res2Net achieves an EER result of 0.47\%. This shows that local attention can capture important information in more detail and reduce the influence of redundant information left over from the network.

	
	Finally, for the setting of the number of channel groups in the SR-LA Res2Net, we believe that n=8 is most appropriate for right in the fake audio detection task, thus achieving state-of-the-art performance. This is because the information exchange at this time is sufficient and the feature representation is reasonable.
	
	Moreover, to verify the effectiveness of our proposed SR-LA Res2Net, we simulate the F0 subband with some recently published advanced networks. Table~\ref{tab:result3_parm} shows the EER and t-DCF of different networks, which demonstrates the differences between different networks when using either the original features or the F0 subband.
	Fig.~\ref{fig:Other} shows the specific EER results of the attack. From Fig.~\ref{fig:abolation} and Fig.~\ref{fig:Other}, we see that the F0 subband feature can achieve excellent performance when paired with other networks, and SR-LA Res2Net can fully access the discriminative information of the F0 subband and perform extremely well in all aspects.

	\vspace{1ex}
	\subsubsection{Effective Generalization Ability of SR-LA Res2Net Architecture}\
	

	\begin{table}[!t]
		\caption{EER and t-DCF of single systems and primary systems based on the top performance of ASVspoof 2019 LA dataset.}	
		(a) Single systems
		\label{tab:result4}
		\renewcommand\arraystretch{1.2}
		\centering
		\begin{tabular}{p{50mm}cc}
			\hline \toprule System & t-DCF & EER\% \\ \hline
			CQCC+GMM (B1) & $0.2316$ & $9.57$ \\
			LFCC+GMM (B2) & $0.2116$ & $8.09$ \\
			LFCC-Siamese CNN \cite{Lei2020-et-al:scheme} & $0.0930$ & $3.79$ \\
			RW-ResNet \cite{Ma-et-al:scheme} & $0.0820$ & $2.98$ \\	
			ACNN \cite{Ling2021} & $0.0510$ & $1.87$ \\
			MCG-Res2Net50 \cite{li2021channel} & $0.0520$ & $1.78$ \\		
			FFT-L-SENet \cite{Zhang-et-al:scheme} & $0.0368$ & $1.14$ \\ 
			AASIST \cite{jung2022aasist} & $0.0347$ & $1.13$ \\ 
			SAMO \cite{ding2023samo} & $0.0356$ & $1.08$ \\ 
			PA-Res2Net \cite{kim2023phase} & $0.0300$ & $1.07$ \\ 
			ECANet\_SD \cite{xue2023learning}  & $0.0295$ & $0.88$ \\ 
			\hline
			\textbf{Ours (single system)} & $\textbf{0.0159}$ & $\textbf{0.47}$ \\

			\hline \toprule
		\end{tabular}
		
		\vspace{3ex}	
		(b) Fusion systems
		\label{tab:resultb}
		\renewcommand\arraystretch{1.2}
		\centering
		\begin{tabular}{p{50mm}cc}
			\hline \toprule System & $\mathrm{t}$-DCF & EER\% \\
			\hline T05 \cite{Todiscoetal} & $0.0069$ & $0.22$ \\
			T45 \cite{Lavrentyeva-et-al:scheme} & $0.0510$ & $1.84$ \\
			T60 \cite{chettri-et-al:scheme} & $0.0755$ & $2.64$ \\
			GMM fusion \cite{Tak-et-al:scheme} & $0.0740$ & $2.92$ \\
			T24 \cite{Todiscoetal} & $0.0953$ & $3.45$ \\
			T50 \cite{Yang-et-al:scheme} & $0.1671$ & $3.56$ \\ \hline
			\textbf{Ours (single system)} & $\textbf{0.0159}$ & $\textbf{0.47}$ \\
			\hline \toprule
		\end{tabular}
	\end{table}
	
	Fig.~\ref{fig:abolation} counts the EER of each attack algorithm for different systems.
	 From Fig.~\ref{fig:abolation}, we can draw the following points. First, when different classifiers capture the details of the feature and then generalize to unknown attacks, their biases will be large. For example, the GMM-based baseline algorithm is very effective against attack algorithms such as A08, but the performance of attack algorithms such as A10, A13, A14, and A17 becomes extremely poor. Even so, it is not our original intention to only effectively target a certain attack algorithm. For example, our proposed SR-LA Res2Net (n=8, F0) can be generalized to each unseen attack more evenly, which is the focus of our work.
	
	Second, it is well known that the A17 algorithm is notorious, and the method was judged to have the highest spoofing ability in the 2018 Speech Transformation Challenge~\cite{kinnunen2018spoofing}. 
	However, our proposed SR-LA Res2Net system can obtain 0.70\% EER on the A17 attack, which is the best performance among all systems.
	We believe that the multi-scale feature representation enables the \acrshort{fsd} system to be generalized to spoofing attacks like A17. the EER results of the ResNet (F0) and Res2Net (F0) systems on A17 are 4.43\% and 3.09\%, respectively, which indicates that multi-scale features can greatly extend the feature receptive field and enhance its generalizability. However, the channel group information of its Res2Net architecture is constantly superimposed, which requires a spatial reconstruction block to reconstruct each channel group information, and the EER result of its SR Res2Net (F0) at A17 is 1.95\%, and the experimental results prove that the spatial reconstruction block can reduce the influence of redundant information. In addition, other systems have poor performance for unseen attacks like A08, A17, and A18, but the SR-LA Res2Net (F0) system achieves high performance in the face of all unseen attacks. This further validates the need to integrate spatial reconstruction block and local attention block in Res2Net, which can greatly improve the generalization ability of the model.

	Third, compared to our proposed SR-LA Res2Net (F0) system, other systems are difficult to pass in some individual deception algorithms. For example, the B1 system achieves an EER result of 26.15\% in the A13 algorithm, and our SR-LA Res2Net (F0) system has an EER of 0.09 for A13, and other systems also performed well. For A18, the SR-LA Res2Net (F0) system leads the way.

	In summary, the strong generalization of the SR-LA Res2Net (F0) system comes from the spatial reconstructed block and the local attention block. By reconstructing the feature space and focusing on local information, it reduces the multi-scale sequelae brought by feature representation, which greatly improves the performance of the \acrshort{fsd} system.

	\begin{table*}[]
		\centering
		\caption {EER and t-DCF results for conventional pitch features and F0 subbands under different models. F0 (3D) indicates that additional pitch features (probability of voicing, pitch value and delta pitch value) are used together.}
		\label{Fig:APF}
	
			\begin{tabular}{ccccc}
				\hline
				& Front-end                  & Res2Net         & LA Res2Net      & SR-LA Res2Net   \\ \hline
				\multirow{7}{*}{EER(\%)} & MFCC                       &        9.06         &     8.26        &          7.77       \\
				& MFCC + F0 (3D)             & 7.92            & 7.72            & 7.39            \\ \cline{2-5} 
				& LFCC                       & 4.92            & 2.28            & 2.05            \\
				& LFCC + F0 (3D)             & 3.76            & 2.42            & 1.99            \\ \cline{2-5} 
				& Low subband                & 1.85           & 2.82            & 2.06            \\
				& Low subband + F0 (3D)      & 1.82            & 1.53            &  1.47               \\ \cline{2-5} 
				& F0 subband (\textbf{Ours}) & \textbf{1.17}   & \textbf{0.80}   & \textbf{0.47}   \\ \hline
				\multirow{7}{*}{t-DCF}  & MFCC                       &   0.2220              &     0.1825            &     0.2146      \\
				& MFCC + F0 (3D)             & 0.1836          & 0.1779          & 0.2312          \\ \cline{2-5} 
				& LFCC                       & 0.1358          & 0.0652          & 0.0552          \\
				& LFCC + F0 (3D)             & 0.0815          & 0.0666          & 0.0538          \\ \cline{2-5} 
				& Low subband                & 0.0510          & 0.0561          & 0.0577          \\
				& Low subband + F0 (3D)      & 0.0597          & 0.0455          &   0.0389              \\ \cline{2-5} 
				&F0 subband (\textbf{Ours})  & \textbf{0.0353} & \textbf{0.0246} & \textbf{0.0159} \\ \hline
		\end{tabular}
	\end{table*}
	
	\begin{table}[!t]
		\caption{EER and t-DCF of single systems and primary systems based on the top performance of ASVspoof 2021 LA dataset.}	
		(a) Single systems
		\label{tab:result5}
		\renewcommand\arraystretch{1.2}
		\centering
		\begin{tabular}{p{50mm}cc}
			\hline \toprule System & t-DCF & EER\% \\ \hline
			
			B03 \cite{Yamagishi} & $0.3445$ & $9.26$ \\
			B04 \cite{Yamagishi} & $0.4257$ & $9.50$ \\ 
			B01 \cite{Yamagishi} & $0.4974$ & $15.62$ \\ 
			B02 \cite{Yamagishi} & $0.5758$ & $19.30$ \\ \hline
			\textbf{Ours (single system)} & $\textbf{0.2642}$ & $\textbf{3.61}$ \\
			\hline \toprule
		\end{tabular}
		
		\vspace{3ex}	
		(b) Fusion systems
		\label{tab:resultb}
		\renewcommand\arraystretch{1.2}
		\centering
		\begin{tabular}{p{50mm}cc}
			\hline \toprule System & t-DCF & EER\% \\ 
			\hline T23 \cite{tomilov21_asvspoof} & $0.2177$ & $1.32$ \\
			T20 \cite{chen21b_asvspoof} & $0.2608$ & $3.21$ \\
			T04 \cite{caceres21_asvspoof} & $0.2747$ & $5.58$ \\
			T06 \cite{kang21b_asvspoof} & $0.2853$ & $5.66$ \\ \hline
			\textbf{Ours (single system)} & $\textbf{0.2642}$ & $\textbf{3.61}$ \\ 
			\hline \toprule
		\end{tabular}
		\vspace{2ex}
	\end{table}

	\vspace{1ex}
	\subsubsection{Comparison with Other Systems}\
	
	\vspace{1ex}
	Table~\ref{tab:result4} shows the results of the eight best-performing single systems, the six main systems, and our best system on the ASVspoof 2019 LA evaluation set. where B1 and B2 are the baseline systems.
	The results of single systems are shown in Table~\ref{tab:result4}a. These systems include some top-performance systems from the ASVspoof 2019 challenge and systems from recently published papers. 
	Table~\ref{tab:result4}b shows the results of the primary systems, where T05, T45, T60, T24, and T50 represent the anonymous identifiers of the teams in the ASVspoof 2019 challenge. These primary systems may contain multiple front-end features and neural network architectures. The GMM fusion system consists of the nonlinear fusion of its six subbands. From the performance comparison of different systems in Table~\ref{tab:result4}, it can be seen that our proposed system achieves state-of-the-art performance in a single system, and also outperforms the second-ranked system on the ASVspoof 2019 LA challenge among the primary systems.

	To the best of our knowledge, among all fusion systems, only the T05 system outperforms us. Here we want to emphasize that the fusion system is obtained by fusing multiple single systems, which means that multiple models need to be trained and finally fused, so that the overall model parameters are huge. Moreover, T05 is a fusion of 7 single systems, including 2 Resnet models, 4 MobileNet models, and 1 DenseNet model, and the final results are obtained by combining the equal weights of these 7 single systems. It can be seen that the T05 system architecture is extremely complex. Therefore, our proposed single system has advantages in terms of performance and network architecture.
	\vspace{1ex}
	\subsubsection{t-SNE Visualization Analysis}\
	
	To visualize the effectiveness of the proposed approach, we also visualize the baseline system and my proposed system using t-SNE~\cite{hinton2008visualizing}, respectively. Both models are trained on the LA dataset in Asvspoof 2019 and take the penultimate layer of the network. As shown in Fig.~\ref{fig:t-sne}, we can see that the real and fake speech of the Low subband and SENet systems are not distinguished, and there are many blue dots embedded inside the red dots. While the true and false speech of the F0 subband and SR-LA Res2Net systems are separated, there is only a little blending at the boundary. The above visualization results further validate our experimental results.

	\subsection{Experimental Results on ASVspoof 2021 LA dataset}
	
	Table~\ref{tab:result5} shows the results of the ASVSpoof 2021 LA Challenge for single and fusion systems.
	Among them, the T23~\cite{tomilov21_asvspoof} system is a fusion of twelve subsystems, including ten MSTFT-LCNN systems, one MSTFT-ResNet system, and one RawNet system, finally fused in the scoring stage by a fine weight assignment; the T20\cite{chen21b_asvspoof} system is a fusion of three subsystems based on ResNet system with equal weight fusion of scoring; the T04\cite{caceres21_asvspoof} system is a scoring fusion of three subsystems, namely LFCC-LCNN, RawNet2, and lightweight TDNN Focal, and all three subsystems use a data enhancement strategy; the T06\cite{kang21b_asvspoof} system is a fusion of eight subsystems, namely an LFCC-LCNN system (baseline), one RawNet2 system (baseline), one LFCC-GMM, four LFCC-SENet systems, and one PSCC-TDNN system, all of which use data enhancement strategies except for the baseline system. B01-B04 are the four baseline systems for the ASVSpoof 2021 LA Challenge. The following points can be observed from the table:
	
	(1) Most of the systems perform data enhancement in the face of transmission interference in the ASVspoof 2021 LA data set, and the performance is poor for the baseline systems that do not perform data enhancement.
	
	(2) Several of the most advanced systems submitted at the ASVspoof 2021 LA challenge are fusion systems, while we can obtain good performance for our single system.
	
	In conclusion, our proposed single system is competitive in the face of both single and fusion systems, which further validates the effectiveness of our proposed method.

	\subsection{Comparison with conventional F0 extraction methods}
	We extracted traditional 3D additional pitch features as auxiliary features through the Kaldi tool, which was used in combination with the 80-dimensional MFCC, 60-dimensional LFCC, and low subband. As shown in Table~\ref{Fig:APF}, 3D denotes the additional three-dimensional pitch features. The results in Table~\ref{Fig:APF} show that the additional pitch features can indeed improve the Kaldi performance of the FSD system, which also verifies that F0 has effective discriminative information. However, the system performance is still limited compared to the F0 subband, which also shows the superiority of the F0 subband features.

	\section{Discussion}

		In this section, we discuss the advantages, shortcomings, and future directions of the proposed method in the current research of  \acrshort{fsd}.

		First, our proposed system has a high-performance advantage, because the F0 subband features have a strong discriminative ability to distinguishing between real and fake speech, and the SR-LA Res2Net can further utilize the multi-scale discriminative information of F0 subband to improve the performance of \acrshort{fsd}. In addition, from the results of the ASVspoof 2021 LA database, it can be found that our proposed single system can still achieve competitive performance even in a fusion system.

		Further, the proposed system also has obvious advantages in real scenarios and practical applications. The first one is to deal with a large amount of data, like about 180,000 entries on the ASVsopoof 2021 LA dataset, which can still be handled efficiently by the proposed system. The second is the challenge of practical deployment, most of the current systems are deployed in automatic speaker verification (ASV), automatic Speech Recognition (ASR), etc., which may require lower computational requirements, and the proposed system is also less than 1MB in model parameters, which is also very convenient to use for ensembling other existing systems. On the other hand, since the system is mostly deployed for ASV and ASR systems, the communication interference encountered in real scenarios is the most common. Thus, the ASVspoof 2021 LA dataset models a large number of communication disturbances, specifically real and spoofing speech transmitted using a variety of telephony systems, including Voice over IP (VoIP) and Public Switched Telephone Network (PSTN).

		Finally, we summarize the future directions of \acrshort{fsd} systems. 1) Improve the generalization ability of the system in the face of unknown attacks; 2) Enhance the robustness of the system in complex scenarios, such as bandwidth noise, communication interference, and other real-world conditions; 3) Explore the study of lightweight of the system, which can have efficient flexibility in specific applications; 4) Explain the decision-making process of the system in depth. The current research on \acrshort{fsd} is not able to do the above points better, which is a future research trend, including the ASVspoof dataset also needs to make more progress for more application scenarios.

	\section{Conclusions}
	
In this paper, we propose an F0 subband with SR-LA Res2Net for \acrshort{fsd}. The F0 distribution of bonafide speech is often difficult to replicate so it is very different from the fake one. Therefore, we think the F0 contains the discriminative information. In addition, to effectively model the F0 subband, we propose a novel SR-LA Res2Net for \acrshort{fsd}.
Specifically, the SR block is designed to eliminate spatial artifacts when information is transmitted between channel groups. The LA block is used to focus on local information. Experimental results on the ASVspoof 2019 LA dataset show that our proposed approach is very effective against unseen spoofing attacks and achieves a minimum t-DCF of 0.0159 and an EER of 0.47\%, which achieves state-of-the-art performance among all single systems. One of the limitations of this work is that we only use the F0 subband for \acrshort{fsd}. The other speech information is abandoned, which may also contain some important discriminative information.
In the future, to make full use of the speech information, we will explore combining the F0 subband with other speech features to further improve the performance of \acrshort{fsd}.

	\section{Acknowledgements}
	
	This work is supported by the {STI 2030—Major Projects (No.2021ZD0201500)}, the National Natural Science Foundation of China (NSFC) (No.62201002), Excellent Youth Foundation of Anhui Scientific Committee (No.208085J05), Special Fund for Key Program of Science and Technology of Anhui Province (No.202203a07020008), the Open Research Projects of Zhejiang Lab (NO.2021KH0AB06).

	
	\bibliographystyle{cas-model2-names}
	
	\bibliography{cas-refs}


\end{document}